\documentclass{article}

\usepackage{mathtools}  
\usepackage{amsmath}
\usepackage{amssymb}  

\usepackage{spconf}

\usepackage[final]{microtype}

\usepackage{graphicx}

\usepackage{booktabs}  

\usepackage[font=small,labelfont=bf]{caption}
\usepackage{subcaption}

\usepackage[hidelinks,pdfpagemode=UseNone]{hyperref}
\usepackage{url}
\usepackage[capitalise]{cleveref}

\usepackage{siunitx}
\DeclareSIUnit{\nounit}{\relax}

\newcommand{\durhm}[2]{\SI{#1}{\hour}\,#2}  

\usepackage[%
  olditem,  
  oldenum   
]{paralist}

\makeatletter
\newcommand*{\transpose}{%
  {\mathpalette\@transpose{}}%
}
\newcommand*{\@transpose}[2]{%
  \raisebox{\depth}{$\m@th#1\intercal$}%
}
\makeatother

\let\norm\undefined
\DeclarePairedDelimiter\norm{\lVert}{\rVert}

\newcommand{\mlabel}[1]{\texttt{\mbox{#1}}}

\title{Semi-supervised Learning for Singing Synthesis Timbre}

\name{Jordi Bonada, Merlijn Blaauw}
\address{%
Music Technology Group, Universitat Pompeu Fabra, Barcelona, Spain\\
{\small\href{mailto:jordi.bonada@upf.edu}{\nolinkurl{jordi.bonada@upf.edu}}, \href{mailto:merlijn.blaauw@upf.edu}{\nolinkurl{merlijn.blaauw@upf.edu}}}%
}

\begin{document}

\maketitle

\begin{abstract}
We propose a semi-supervised singing synthesizer, which is able to learn new voices from audio data only, without any annotations such as phonetic segmentation. Our system is an encoder-decoder model with two encoders, linguistic and acoustic, and one (acoustic) decoder. In a first step, the system is trained in a supervised manner, using a labelled multi-singer dataset. Here, we ensure that the embeddings produced by both encoders are similar, so that we can later use the model with either acoustic or linguistic input features. To learn a new voice in an unsupervised manner, the pretrained acoustic encoder is used to train a decoder for the target singer. Finally, at inference, the pretrained linguistic encoder is used together with the decoder of the new voice, to produce acoustic features from linguistic input. We evaluate our system with a listening test and show that the results are comparable to those obtained with an equivalent supervised approach.
\end{abstract}



\section{Introduction}\label{sec:intro}

Singing synthesis has recently seen a notable uptick in research activity, and, inspired by modern deep learning techniques developed for text-to-speech (TTS), great strides have been made, e.g.~\cite{BlaauwM2017_NPSS_MDPI,LeeJ2019_KoreanSS,BlaauwM2020_TransformerSVS,AngeliniO2019_AmazonSVS,GuY2020_ByteDanceSVS,WuJ2020_AdversarialSeq2SeqSS,LuP2020_XiaoiceSing,ChenJ2020_HiFiSinger}. To create a new voice for these models, generally a supervised approach is used, meaning that besides recordings of the target singer, phonetic segmentation or a reasonably well-aligned score with lyrics is needed. These annotations generally have to be done manually, or at least corrected by hand, causing them to become a bottleneck when creating new voices. Having a wide array of voices in singing synthesis is generally desirable as most applications tend to be creative (unlike TTS). Other use cases such a choir singing also require a great number of voices.

One approach to singing synthesis is to break up the process into two steps; first a pitch model generates an F0 curve given a musical score (with lyrics), and then a timbre model produces the final acoustic rendition given F0 and a timed phonetic sequence as input. While it is possible to combine these two steps, e.g.~\cite{LuP2020_XiaoiceSing,ChenJ2020_HiFiSinger}, in this work we use the two-step approach and focus on the timbre model only. The main reason for this is that we consider both tasks to have notably different requirements and constraints, and thus prefer to focus on each step individually. We will tackle the task of semi-supervised training of a pitch model in a separate paper.

The principal contributions of this paper are:
\begin{inparaenum}
  \item A semi-supervised method for learning the timbre of new voices from audio data only.
  \item A method for controlling a single decoder with embeddings derived from either acoustic or linguistic features.
  \item Distinction between long and short scopes in the decoder for tackling the phonetic context of long vowels in singing while producing detailed acoustic features.
  \item Evaluation of the proposed system.
\end{inparaenum}

\section{Relation to prior work}\label{sec:prior_work}



While semi-supervised singing synthesis has not been widely studied, the task of non-parallel voice conversion is closely related. Most of these approaches try to extract the linguistic content from a given audio signal, independently of factors such as speaker identity, pitch, loudness, and so on. The majority of approaches are based on the autoencoder, with several reoccurring themes that ensure only linguistic content is encoded. One such theme is to use information-restrictive bottleneck, e.g. by temporal downsampling \cite{QianK2019_AutoVC}, carefully selected dimensionality \cite{QianK2019_AutoVC}, variational regularization \cite{LuoYJ2020_VAESingingVC}, or vector quantization \cite{VanDenOordA2017_VQVAE}. This is often combined with a decoder conditioned on non-linguistic factors, such as speaker embedding \cite{QianK2019_AutoVC}, or F0 \cite{QianK2020_F0ConsistentAutoVC,PolyakA2020_UnsupervisedCrossDomainSingingVC}. Data augmentation can be used to make the encoder more invariant to aspects we do not wish to encode, e.g. using pitch shifting or time stretching \cite{QianK2020_F0ConsistentAutoVC,QianK2020_TrippleAutoVC}. The negative gradient of an auxiliary classifier can be used to reduce undesirable information in the bottleneck, e.g. a speaker classifier to reduce information related to speaker identity \cite{NachmaniE2019_UnsupervisedSingingVC}. Cycle-consistency is another common theme, which relies on the fact that conversion to another speaker identity and back should result in a similar output \cite{KanekoT2018_CycleGANVC,KanekoT2019_CycleGANVC2,KameokaH2018_StarGANVC,KanekoT2019_StarGANVC2}. This technique is often combined with adversarial training, which is also applied in a number of other approaches \cite{PolyakA2020_UnsupervisedCrossDomainSingingVC}. Backtranslation techniques can be used to generate parallel training data \cite{NachmaniE2019_UnsupervisedSingingVC,PolyakA2020_UnsupervisedCrossDomainSingingVC}. Finally, using a phonetic recognizer trained on audio with transcription can aid extracting content features, albeit in a more supervised manner \cite{PolyakA2020_UnsupervisedCrossDomainSingingVC,SunL2016_VoiceConversionPPG}.


In singing synthesis, several works aim to go towards a reduction in the burden of dataset annotation. In particular, sequence-to-sequence models generally avoid the need of detailed phonetic segmentation, but do require a fairly well aligned musical score with lyrics \cite{LeeJ2019_KoreanSS,BlaauwM2020_TransformerSVS,AngeliniO2019_AmazonSVS,GuY2020_ByteDanceSVS,WuJ2020_AdversarialSeq2SeqSS,LuP2020_XiaoiceSing,ChenJ2020_HiFiSinger}. Similarly voice cloning techniques require only a small amount of training data with phonetic segmentation for the target voice (e.g. \SI{3}{\minute} versus an hour or more) \cite{BlaauwM2019_VoiceCloningSS}. However, this limited data regime may ultimately affect sound quality, and still requires some annotation effort. Finally, some work has been done to generate voices from data mined from the web (audio and lyrics) in a completely automatic manner by aligning lyrics to audio \cite{RenY2020_DeepSinger}.

\section{Proposed system}\label{sec:proposed_system}

\begin{figure*}[htb]  
  \centering
  \includegraphics[width=17.8cm]{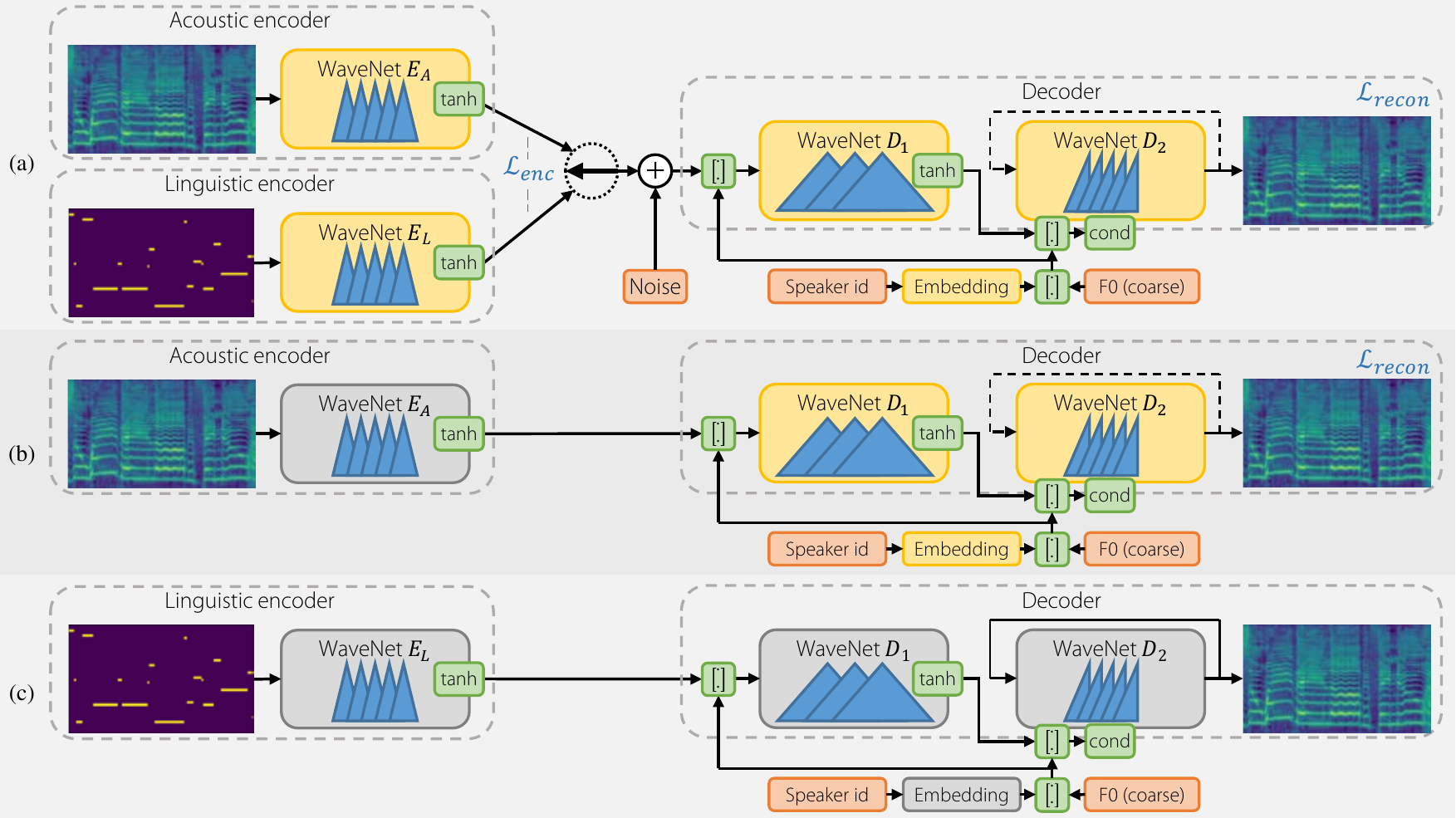}
  \begin{subfigure}{0em}
    \phantomsubcaption{}
     \label{fig:model_architecture:a}
  \end{subfigure}
  \begin{subfigure}{0em}
    \phantomsubcaption{}
     \label{fig:model_architecture:b}
  \end{subfigure}
  \begin{subfigure}{0em}
    \phantomsubcaption{}
     \label{fig:model_architecture:c}
  \end{subfigure}
  \caption{A diagram of the model architecture in three different phases: (a) Training the encoder-decoder from annotated audio (supervised). (b) Training the decoder from audio (unsupervised). (c) Inference from linguistic features. Gray colored modules indicate their weights are kept fixed. The shape of the triangles in the WaveNet blocks represents the size of the receptive field and whether it is causal. A dashed autoregressive connection in the $D_2$ WaveNet block indicates teacher forced training with additive noise to avoid overfitting, while a solid connection indicates true autoregressive inference.}
  \label{fig:model_architecture}
\end{figure*}

As illustrated in \cref{fig:model_architecture}, our system is an encoder-decoder model with two encoders (linguistic and acoustic) and one decoder (acoustic).
In a first step (\cref{fig:model_architecture:a}), the system is trained in a supervised manner, using a labelled multi-singer dataset.
Our aim is that the system can learn similar embeddings from either acoustic or linguistic data obtained from a singing voice. This way, we can interchange the encoders depending on the task.
Hence, to learn a new voice in an unsupervised manner, the pretrained acoustic encoder is used to train a decoder for the target singer using only acoustic data (\cref{fig:model_architecture:b}). Next, at inference, the pretrained linguistic encoder is used together with the decoder of the new voice, to produce acoustic features from linguistic input (\cref{fig:model_architecture:c}). 

Both encoders and decoders are based on the same building block; a WaveNet architecture \cite{VanDenOordA2016_WaveNet} consisting of a set of dilated 1d-convolutional layers featuring gated units, residual shortcut connections and skips, and with the skip sum feeding an output stack of two convolutional layers. 

\subsection{Encoder}
Our system combines an acoustic and a linguistic encoder that are trained to produce similar embeddings when computed from an annotated singing voice. Both encoders are non-causal and share the same structure and hyper-parameters.
The acoustic encoder, $E_A$, takes as input a mel-spectrogram, while the linguistic encoder, $E_L$, takes as input a phonetic segmentation, encoded as a frame-wise sequence of one-hot phoneme encodings.
The encoders have a rather short receptive field of few hundred milliseconds. The reason is that our aim is not to capture large-scale variations, but to focus on a rather short scale, and get embeddings closer to what might be a phonetic transcription. Modeling longer phonetic sequences or clusters is a task left for the decoder. 

\subsubsection{Bottleneck and stochastic switch}
For favouring similar embeddings from acoustic and linguistic data, inspired by \cite{VanV2017_GoogleRandomSwitch}, we randomly switch between acoustic and linguistic encoders during training. Additionally, we add noise to the encoder output (after the non-linearity) as a bottleneck, with the idea that the encoder should not encode mel-spectrogram details, also to encourage more stable embeddings along phonemes. A similar bottleneck is used in \cite{SalakhutdinovR2009_SemanticHashing,KaiserL2018_ImprovedSemanticHashing}, where additive noise is added to the embeddings, however in this case before the non-linearity for saturating it and producing more binary-like embeddings, which is not our goal. 

\subsection{Decoder}
The decoder is divided into a long-scope and a short-scope networks. The first one, $D_1$, focuses on capturing the encoded timbre variations at a large scale of a few seconds, while the second one, $D_2$, focuses on producing a detailed timbre output. 
Vowels can last for some seconds in singing, so capturing the phonetic context requires a receptive field as large as that of $D_1$.
$D_1$ is non-causal and receives the encoder output concatenated with F0 and speaker embedding. 
$D_2$ is an autoregressive convolutional network with a short receptive field (less than \SI{200}{\milli\second}) that produces the final mel-spectrogram. Its architecture combines autoregressive and control inputs as in \cite{BlaauwM2017_NPSS_MDPI}. In our case, the input is the previous mel-spectrogram, and the control input is the $D_1$ output concatenated with F0 and speaker embeddings.

\subsection{Training loss}
The training loss, $\mathcal{L}$, is a weighted sum of two components,
\begin{equation}
\mathcal{L} = \lambda_{recon}\mathcal{L}_{recon} + \lambda_{enc}\mathcal{L}_{enc}.
\end{equation}
Here, $\mathcal{L}_{enc}$ is the (squared) L2 distance between acoustic and linguistic embeddings, produced by acoustic and linguistic encoders, $E_A(\cdot)$ and $E_L(\cdot)$, from aligned acoustic and linguistic input features, $x$ and $y$ respectively,
\begin{equation}
\mathcal{L}_{enc} = \norm{E_A(x) - E_L(y)}^2_2.
\end{equation}
We do this in order to constraint the system to produce similar embeddings either type of input feature.
The reconstruction loss, $\mathcal{L}_{recon}$, is the (squared) L2 distance between the output of the decoder, $\hat{x}$, and the groundtruth acoustic features, $x$,
\begin{equation}
\mathcal{L}_{recon} = \norm{x - \hat{x}}^2_2.
\end{equation}
To compute the decoder output, $\hat{x}$, we first compute its input embedding as a random switch between the acoustic and linguistic embeddings,
\begin{equation}
e = k E_A(x) + (1 - k)E_L(y),
\end{equation}
where $k \sim Bern$. Then random noise, $\epsilon_1 \sim \mathcal{N}(0, \sigma_1)$, is added to the selected embedding, we concatenate with control features, $c$, derived from speaker embedding and F0, and feed the result to the first (non-causal) decoder network, $D_1(\cdot)$. The output of this first decoder network is then concatenated with $c$ and used as a control input on which the second, autoregressive decoder network, $D_2(\cdot,\cdot)$, is conditioned. During the training of $D_2$ we use teacher forcing, where past timesteps are groundtruth acoustic features with added noise, $\epsilon_2 \sim \mathcal{N}(0, \sigma_2)$, to reduce overfitting. Thus, the resulting computation becomes,
\begin{equation}
\hat{x}_i = D_2(x_{<i} + \epsilon_2, [D_1([e + \epsilon_1, c]), c]).
\end{equation}
The constants, $\lambda_{recon}$, $\lambda_{enc}$, $\sigma_1$, and $\sigma_2$, are defined in \cref{ssec:system_settings}.

\subsection{Data-augmentation for improved invariance}
When training our system, we randomly transpose the pitch of the acoustic input without informing the acoustic encoder of the actual transposition factor. This transposition is performed by combining resampling of the input audio signal with time-scaling of the corresponding mel-spectrogram. This transformation modifies the pitch of the signal but also linearly scales the timbre in frequency. Since both acoustic and linguistic embeddings are constrained to be similar, and the linguistic encoder does not depend on the pitch transposition factor, then transposing the acoustic input helps to produce a more speaker and pitch independent embedding.

\section{Experiments}\label{sec:experiments}

\subsection{Datasets}
For the experiments in this work, we use two proprietary datasets. For training the encoders we use a dataset of 7 native English singers (\durhm{5}{47}), which we label \mlabel{DAT7}, with approximately 10 songs per singer (one used for validation, the rest for training). The audio files were phonetically segmented with manual corrections. For training the decoders we use a dataset of 41 pop songs performed by a professional English male singer, labelled \mlabel{DAT1}. From this dataset 38 songs were used for training (\durhm{2}{7} total), 3 for validation (\SI{10}{\minute}). 

\subsection{System settings}\label{ssec:system_settings}
Our proposed system uses 100-dimensional mel-spectrogram acoustic features, extracted with a \SI{45}{\milli\second} window and a \SI{5}{\milli\second} hop time, and computed between \SIrange{10}{15200}{\hertz}. Linguistic features are computed with 1-hot encodings using 43 phonetic symbols. The encoder networks $E_A$ and $E_L$ have 9 non-causal 1d-convolutional layers (3x1), with dilation factors \{1,2,4,1,2,4,1,2,4\}, and 70 residual channels. A leaky ReLU activation follows the skip sum, and then the output stack has 2 convolutional layers with 120 channels, and leaky ReLU and tanh activations respectively. The first block of the decoder $D_1$ contains 10 non-causal 1d-convolutional layers (3x1), dilation factors \{1,2,4,1,2,4,1,2,4,1\}, and 70 residual channels. The output stack has the same configuration as for the encoders. Finally, the second block of the decoder $D_2$ has 8 causal 1-d convolutional layers (2x1), 200 residual channels, and dilation factors \{1,2,4,8,16,1,2,4\}. The skip sum is followed by a leaky ReLU activation. The first convolution in the output stack has 200 channels and leaky ReLU activation, and the second one directly predicts the output mel-spectrogram. All leaky ReLU activations use $\alpha=0.2$.
Speaker embeddings are computed as 1-hot encodings followed by a 1x1 convolution with 16 channels. 

We use the Adam optimizer with $\beta_1=0.9$, $\beta_2=0.999$, $\epsilon=\num{1e-8}$, and a batch size of 12. We follow the learning rate schedule from \cite{VaswaniA2017_AttentionIsAllYouNeed}, with a 700 step warm-up, a base learning rate of \num{5e-4}, a decay rate of \num{0.15} every \num{10000} steps.
The objective that we optimize is an (squared) L2 loss between output and target features. When training encoders, we use an additional (squared) L2 loss between acoustic and linguistic encoded features, with a weighting of $\lambda_{enc}=0.2$ and $\lambda_{recon}=1$. In addition, we add normal noise with $\sigma_1=0.3$ to the embeddings, and with $\sigma_2=0.2$ to the $D_2$ input. Each sample in the minibatch produces a valid output length of \SI{1.5}{\second} (excluding system receptive field).

\subsection{Evaluation}
We compare our proposed semi-supervised model to a similar supervised model, being the only difference between both systems that the supervised model does not have an acoustic encoder. Thus, it learns to predict acoustic features from the input linguistic features using an annotated dataset of the target singer. For the semi-supervised case, we first train the encoder-decoder with \mlabel{DAT7} (as in \cref{fig:model_architecture:a}), and next retrain the decoder with \mlabel{DAT1} (as in \cref{fig:model_architecture:b}). For the supervised model, we train the encoder-decoder directly with \mlabel{DAT1}, as in \cref{fig:model_architecture:a} but without the acoustic encoder.

Additionally, we also evaluate the case of using only a small dataset of training data for the target voice (\SI{3}{\minute}), labelled \mlabel{DAT1-C}. We compare our proposed semi-supervised cloning approach to a supervised approach. For the semi-supervised case, we first train the encoder-decoder with \mlabel{DAT7}, and next we fine-tune the model with \mlabel{DAT1-C} for few thousand updates without using dataset annotations (as in \cref{fig:model_architecture:b}). For the supervised cloning case, we first train the supervised encoder-decoder with \mlabel{DAT7}, and then fine-tune the model with annotated \mlabel{DAT1-C} for a few thousand updates.

We ran a MOS listening test with 12 participants, which each rated a random subset of 12 out of 24 phrases. Per test 6 stimuli were presented; the 4 systems mentioned previously, and visible and hidden references consisting of a re-synthesis of the target recording. All systems were presented and rated together to encourage a comparison between them. Audio was generated with a mel-spectrogram driven neural vocoder \cite{TamamoriA2017_WaveNetVocoder} trained with \mlabel{DAT1}. 
Sound examples are available online\footnote{\url{https://mtg.github.io/singing-synthesis-demos/semisupervised/}}.

\begin{table}
  \centering
  \caption{Mean Opinion Score (MOS) ratings on a 1--5 scale with their respective 95\% confidence intervals.}
  \label{tab:mos}
  \begin{tabular}[b]{@{}lcc@{}}
\toprule
\textbf{System}         & \textbf{Mean Opinion Score} \\
\midrule
Hidden reference        &         4.80 $\pm$ 0.05     \\
\textbf{Supervised}     & \textbf{3.42 $\pm$ 0.12}    \\
Semi-supervised         &         3.37 $\pm$ 0.11     \\
Supervised cloning      &         2.66 $\pm$ 0.12     \\
Semi-supervised cloning &         2.80 $\pm$ 0.11     \\
\bottomrule
\end{tabular}

\end{table}

The results of our listening tests are shown in \cref{tab:mos}. We can see that the semi-supervised and the supervised systems performs similarly, without a very significance difference. Both systems outperform the cloning approaches, probably due to the small amount of target data available for those. Finally, the semi-supervised cloning system is rated slightly better than the supervised one.

\section{Conclusions}\label{sec:conclusions}

In this work we have proposed a semi-supervised method for learning a new voice timbre model from a singing dataset without any annotations. Our system produces a synthetic acoustic rendition given F0 and a timed phonetic sequence as input. According to our evaluation results, our proposed system performs similarly when compared to an equivalent supervised system using manually corrected annotations. This means that we can effectively reduce the effort to learn a new voice, by removing the dataset annotation task, and without significantly degrading the synthesis quality.
This method could be very useful in the context of choir singing, allowing to model many singers without the dataset annotation burden.
Also we showed that the proposed method performs acceptably in low-resource scenarios, where we only have access to a small amount of a capella audio material. 

At inference, we can produce acoustic features from linguistic or acoustic inputs. In some informal experiments, we observed that our approach can be effectively used as a voice conversion system when controlled by acoustic inputs, performing promisingly in cross-lingual scenarios.

\section{Acknowledgments}

This work was funded by TROMPA H2020 No 770376.

\clearpage


\bibliographystyle{IEEEbib-abbrev}  
{\small\bibliography{semisupervised_timbre}}

\end{document}